\documentclass[
 reprint,
 superscriptaddress,
 amsmath,amssymb,
 aps,
 prb,
longbibliography
]{revtex4-2}

\usepackage[final]{graphicx}
\usepackage{dcolumn}
\usepackage{bm}
\usepackage{xcolor}
\usepackage{color,soul}
\usepackage{algorithm}
\usepackage{algpseudocode}
\usepackage{url} 
\usepackage{makecell}
\usepackage{multirow}
\usepackage{booktabs}
\usepackage{subfig}
\usepackage{changes}
\usepackage[normalem]{ulem}
\usepackage{cleveref}

\def\vii{{\bm v}}

\def\BBb{{ B}}
\def\BBbold{{\bm B}}
\def\ace{\varphi}

\def\cBB{c} 
\def\calF{\mathcal{F}}

\def\dev{\mathrm{dev}}

\newcommand{\PM}{{\texttt{\detokenize{pacemaker}} }}

\newcommand{\LMPS}{{\texttt{\detokenize{LAMMPS}} }}

\date{\today}

\begin{document}
\title{Active learning strategies for atomic cluster expansion models}

\author{Yury Lysogorskiy} 
\email[]{yury.lysogorskiy@rub.de}
\affiliation{ICAMS, Ruhr-Universit\"at Bochum, Bochum, Germany}

\author{Anton Bochkarev}
\affiliation{ICAMS, Ruhr-Universit\"at Bochum, Bochum, Germany}

\author{Matous Mrovec}
\affiliation{ICAMS, Ruhr-Universit\"at Bochum, Bochum, Germany}

\author{Ralf Drautz} 
\affiliation{ICAMS, Ruhr-Universit\"at Bochum, Bochum, Germany}

\date{\today}

\begin{abstract}

The atomic cluster expansion (ACE) was proposed recently as a new class of data-driven interatomic potentials with a formally complete basis set.
Since the development of any interatomic potential requires a careful selection of training data and thorough validation, an automation of the construction of the training dataset as well as an indication of a model's uncertainty are highly desirable. In this work, we compare the performance of two approaches for uncertainty indication of ACE models based on the D-optimality criterion and ensemble learning.
While both approaches show comparable predictions, the extrapolation grade based on the D-optimality (MaxVol algorithm) is more computationally efficient.
In addition, the extrapolation grade indicator enables an active exploration of new structures, opening the way to the automated discovery of rare-event configurations.
We demonstrate that active learning is also applicable to explore local atomic environments from large-scale MD simulations.

\end{abstract}
\maketitle

\section{Introduction}

Machine-learned interatomic potentials (MLIPs) have become significantly more accurate than classical interatomic potentials for many material systems~\cite{bartok2015g,thompson2015spectral, shapeev2016moment,deringer2019machine,zhang2020dpgen, cheng2020evidence, smith2021automated, schran2021machine,batzner20223, Qamar2022, sauceda2022bigdml, kocer2022neural}. The accuracy of MLIPs is achieved by employing flexible functional representations that interpolate reference training data with only very small errors. However, MLIPs can have thousands or even millions of trainable parameters in nonlinear functions, which makes it difficult to estimate extrapolation errors for atomic configurations that were not part of the training data. 

The overall transferability of MLIPs can be assessed using similar validation procedures as for classical interatomic potentials. Ideally, the validation tests should be complementary to the training data, but their selection is often a matter of taste and experience of the developer. It is therefore highly desirable to implement standardized procedures to determine extrapolation errors as part of any MLIP development. Only such measures can ascertain both an unbiased assessment of the transferability and an efficient sampling of unseen atomic configurations.

Here, we introduce an automatized extrapolation evaluation for the atomic cluster expansion (ACE) \cite{Drautz19, Drautz20, dusson2019atomic}. ACE is different from most MLIPs in that it provides a complete set of basis functions that span the space of atomic environments. ACE formulations can have both linear and mildly non-linear characters, and may be understood as systematic, physically-based extensions of classical potentials \cite{Drautz19, Bochkarev2022, Qamar2022}. This is in contrast to highly non-linear MLIPs based on neural networks that lack a clear relation to underlying physical principles.

There exist several approaches that are used for uncertainty indication of MLIPs. The simplest ones include  ensemble learning, such as the query-by-committee~\cite{seung1992proceedings}, where several models are trained in parallel, for example, using different weighting strategies or different reference data. The variance is then estimated from the differences in predictions by these models \cite{behler2014representing, behler2015constructing, gastegger2017machine, zhang2020dpgen}. Bayesian methods often employ Gaussian process regression (GPR), which allows the prediction of variance based on local training data density~\cite{bartok2015g, guan2018construction, sivaraman2020machine, vandermause2020fly}.  

Recently, a combination of Bayesian linear regression (BLR) and ensemble sampling has been adopted to estimate the inference and uncertainty of linear ACE models~\cite{van2022hyperactive}. 
However, the uncertainty associated with the trainable parameters of BLR vanishes in the theoretical limit of infinite number of training samples,    
leaving only a constant term representing the noise in the data~\cite{bishop2006pattern}.  
This can potentially limit the sensitivity of BLR for uncertainty quantification for large datasets. 

Different from GPR and ensemble learning are approaches based on the D-optimality criterion~\cite{podryabinkin2017active, gubaev2019accelerating}. The so-called  extrapolation grade enables to determine whether a given atomic configuration can be regarded as an interpolation or extrapolation of the reference training data. 
This approach not only eliminates the necessity of training multiple models, but it also provides the possibility to select more representative training data.

The aim of this work is to assess both ensemble learning and D-optimality approaches in the context of ACE. We show how intermediate ACE parameterizations can be used to discover atomic configurations with large predicted variances or large extrapolation grades. These configurations are then evaluated using reference electronic structure calculations and added to the training data. This leads to a systematic step-by-step improvement of subsequent parameterizations. 

Our extrapolation grade implementation is available in \LMPS~\cite{LAMMPS} and can be used in two main modes. First, by monitoring the extrapolation grade during a simulation run, the application range of ACE may be demarcated by identifying atomic environments that are too different from the training data and therefore possibly not described reliably. Second, by continuous updating of training data with atomic environments that display large extrapolation grades, different active learning (AL) strategies may be used for automated training. We believe that such autonomous AL schemes based on the self-discovery of deficiencies are crucial for automatizing and speed-up of MLIP construction, and should be widely adopted to increase the reliability of MLIP predictions in general.

The paper is structured as follows. In Sec.~\ref{sec:ACE}, we give a short summary of ACE. In Sec.~\ref{sec:uncertain}, we introduce the ensemble learning and D-optimality criteria in the context of ACE. This is followed by application and comparison of both criteria for several representative examples in Sec. \ref{sec:application}. The application of active learning strategies is explored  in Sec.~\ref{sec:AL}  and conclusions are provided in Sec.~\ref{sec:conclusion}.

\section{Atomic cluster expansion}
\label{sec:ACE}

The atomic cluster expansion provides a complete and hierarchical set of basis functions that spans the space of local atomic environments. 
We summarize here only the essentials of ACE relevant to AL.
Details on ACE and its implementation and parameterization can be found in Refs.~\cite{Drautz19, Drautz20, dusson2019atomic,  lysogorskiy2021, Bochkarev2022}.

An atomic property $\ace_i^{(p)}$, which is a function of the local atomic environment of atom $i$, is expanded as
\begin{equation}
\ace_i^{(p)} = \sum_{\vii}^{n_{\vii}} \cBB_{\vii}^{(p)} \BBb_{i \vii} \,, \label{eq:ACEproperty} 
\end{equation}
with expansion coefficients $\cBB_{\vii}^{(p)}$, and $n_{\vii}$ basis functions $\BBb_{i \vii}$ with multi-indices ${\vii}$.  The basis functions fulfill fundamental translation, rotation, inversion and permutation (TRIP) invariances for the representation of scalar variables, or equivariances for the expansion of vectorial or tensorial quantities. 

In the simplest case of only one atomic property, the energy of atom $i$ is evaluated linearly as
\begin{equation}
   \label{eq:ACEenergylinear} 
    E_i = \ace_i^{(1)} \,.
\end{equation}
For more properties, the energy is written as
\begin{equation}
E_i = \calF(\ace_i^{(1)},  \ace_i^{(2)},  \dots, \ace_i^{(P)}) \,, \label{eq:ACEenergy} 
\end{equation}
where $\calF$ is in general a non-linear function. Most ACE parameterizations to date have employed either a single atomic property or two atomic properties with a Finnis-Sinclair square-root embedding,
\begin{equation}
E_i =  \ace_i^{(1)} + \sqrt{\ace_i^{(2)}}  \,. \label{eq:EFS}
\end{equation}

\section{Uncertainty indication methods}
\label{sec:uncertain}

\subsection{Ensemble-based learning}
\label{sec:uncertainty_methods:QBC}

Nonlinear models frequently have loss functions with multiple minima at comparable losses. 
In addition, the loss function is affected by the reference data and weighting schemes. 
Variations in both reference data and weighting may therefore result in different models of comparable quality. 
By training multiple models and reducing the correlation between the parameters in the models, e.g., by using random initialization of parameters or random sub-sampling of reference training dataset, one can generate an ensemble of models. 
The prediction of the ensemble and its uncertainty are then obtained by statistical aggregation~\cite{krogh1995neuralensemble}, using either mean and standard deviation~\cite{behler2015constructing} or maximum deviation~\cite{zhang2020dpgen} of the individual predictions of the ensemble models on the same input.

The main advantages of ensemble learning are their conceptual simplicity and straightforward implementation.  
One of the limiting requirement is a multi-minima loss function, which mostly occurs naturally in non-linear models.
For linear models, ensembles may be generated as well by varying the reference data and weighting schemes.
Generally, the evaluation time scales linearly  with the ensemble size, but speed-ups are possible in practice, for example, by precomputing atomic descriptors. 
Nevertheless, multiple retraining of models using large datasets may still be time consuming.
It should be also noted that ensembles of neural-network potentials are often overconfident, underestimating the uncertainty of the model~\cite{kahle2022quality}.

In this work, we define the maximum deviations of energies and forces similarly to Ref.~\onlinecite{zhang2020dpgen} as,
\begin{eqnarray}
    \label{eq:devEF}
    \mathrm{dev}(E) &=& \max_k |E^{k} - \left\langle E \right\rangle|\,,\\
    \mathrm{dev}(F_i) &=& \max_k |\mathbf{F}_{i}^{k} - \left\langle\mathbf{F}_{i}\right\rangle|,
\end{eqnarray}
where $k=1 \ldots N_\textrm{ens}$ indexes the models in the ensemble, $E^{k}$ is the energy predicted by model $k$, and $\left\langle E\right\rangle$ the ensemble average of the energy. Even though it may seem that deviations of individual atomic energies could be used as well, they do not correlate with the total energy error (see Appendix~\ref{app:atomicdev}).  The force on atom $i$ in ensemble $k$ is given by $\mathbf{F}_{i}^{k}$ while $\left\langle\mathbf{F}_{i}\right\rangle$ is the ensemble force average.
Apart from this, one also has to choose threshold  values for the uncertainty indicator that separate the interpolation and extrapolation regimes.
We examined several strategies for obtaining ensembles of ACE models as well as for the selection of threshold values. 
Details about these procedures are given in Appendix \ref{app:ens_randomization}.

\subsection{D-optimality criterion and MaxVol algorithm}
\label{sec:uncertainty_methods:maxvol}
The D-optimality criterion aims to select an optimal subset from an over-determined linear problem~\cite{settles2009active} and to quantify the extrapolation grade of a new sample~\cite{goreinov2010find, fonarev2016efficient}.
In our analysis, we employ the MaxVol algorithm~\cite{Mikhalev2016},
which was applied in the context of interatomic potentials for the first time by Podryabinkin and Shapeev~\cite{podryabinkin2017active}. 
Our approach is similar to their query strategy QS$_3$, which is built around the description of local atomic environment instead of the whole structure. 
By linearization about optimum parameter values, the MaxVol algorithm was later extended to nonlinear models~\cite{gubaev2019accelerating}. 

We start from a reference dataset that contains an atom of chemical species $\mu$ in $N_{\mu}$ atomic environments. For each atom $i$ with its corresponding environment, we form a vector composed of $n_{\vii}$ ACE basis functions,
\begin{equation}
\label{eq:B_linear}
    \BBbold_i = \left( \BBb_{i 1}, \BBb_{i 2}, \dots,  \BBb_{i n_{\vii}} \right) \,,
\end{equation}
where the index $i = 1, \dots, N_{\mu}$.
These vectors can be written in a matrix form as
\begin{equation}
\label{eq:Bl}
    \hat{B}_{\mu} = \begin{bmatrix} 
    \BBb_{1 1} &  \dots & \BBb_{1 n_{\vii}} \\
    \vdots & \ddots & \vdots \\
    \BBb_{N_{\mu} 1} &  \dots & \BBb_{N_{\mu} n_{\vii}}
    \end{bmatrix}.
\end{equation}
Typically, the number of atomic environments is much larger than the number of basis functions, i.e., $N_{\mu} \gg n_{\vii}$.
From the $N_{\mu}$ vectors that $\hat{B}_{\mu}$ contains, at most $n_{\vii}$ are linearly independent.
Using the MaxVol algorithm~\cite{goreinov2010find,fonarev2016efficient,Mikhalev2016}, we select from $\hat{B}_{\mu}$ a subset of vectors $\BBbold^A_{1}, \BBbold^A_{2}, \dots, \BBbold^A_{n_{\vii}}$ that span the largest volume, i.e., that have the largest absolute value of the determinant. These vectors form the so-called \emph{active set} matrix
\begin{equation}
\label{eq:Al}
    \hat{A}_{\mu} = \begin{bmatrix} 
    \BBb^A_{1 1} &  \dots & \BBb^A_{1 n_{\vii}} \\
    \vdots & \ddots & \vdots \\
    \BBb^A_{n_{\vii} 1} &  \dots & \BBb^A_{n_{\vii} n_{\vii}}
    \end{bmatrix},
\end{equation}
This implies that all vectors $\BBbold_i$ ($i = 1, \dots, N_{\mu}$) can be represented as linear combinations of the vectors in the active set
\begin{equation}
  \BBbold_i = \sum_{k = 1}^{n_{\vii}} \gamma^{(i)}_{k} \BBbold^A_{k} \,, 
\end{equation}
with
\begin{equation}
    | \gamma^{(i)}_{k}| \leq 1.
\end{equation}
The maximum of the absolute values $| \gamma^{(i)}_{k}|$ for a particular atom $i$ of type $\mu$ defines the extrapolation grade
\begin{equation}
\gamma_{i} = \max_k( | \gamma^{(i)}_{k}|) =\max \left| \BBbold_{i } \cdot \hat{A}_{\mu}^{-1}\right| \,. \label{eq:gamma1}
\end{equation}
It should be noted that multiplication of a particular basis function by a constant for all atomic environments leaves the extrapolation grade unchanged, so that the extrapolation grade is scale invariant (see below). 
If a new atomic environment $i^*$ is encountered, for instance, during a simulation, it is considered as an interpolation if $\gamma_{i^*} \leq 1$ and as an extrapolation if $\gamma_{i^*} > 1$.

In practice, one needs to construct first the active set $\hat{A}_{\mu}$ from a large number of atomic environments present in the training set. Every time a new atomic environment $i^*$ with $\gamma_{i^*} > 1$ is detected in simulations, the active set can be iteratively updated to include the new environment. The MaxVol algorithm~\cite{Mikhalev2016,goreinov2010find, fonarev2016efficient} enables to achieve this in a computationally efficient way, as only simple vector-matrix multiplications, which scale quadratically with the number of basis function, are required.

The evaluation of the extrapolation grade according to Eq.(\ref{eq:gamma1}) is based on the assumption of linear ACE model [cf. Eq.(\ref{eq:ACEproperty}) and (\ref{eq:ACEenergylinear})].
In non-linear ACE formulations, [cf. Eq.(\ref{eq:ACEenergy}) and (\ref{eq:EFS})], one may argue that the D-optimality criterion  should be reformulated. 
The generalized D-optimality criterion \cite{gubaev2019accelerating} linearizes the non-linear equation around the optimized or initial values of trainable parameters $\cBB_{\vii}^{(p)}$ for all properties
\begin{equation}
\BBbold_i = \left( \dfrac{\partial E_i}{\partial \cBB_1}, \dfrac{\partial E_i}{\partial \cBB_2}, \dots,  \dfrac{\partial E_i}{\partial \cBB_{p_{\mu}}} \right) \,, \label{eq:Bnl}
\end{equation}
where the number of expansion coefficients $p_{\mu}$ is given by $p_{\mu} = P\, n_{\vii}$ for $P$ atomic properties.

For the Finnis-Sinclair embedding in Eq.(\ref{eq:EFS}), the left half of the $\hat{B}_{\mu}$ matrix (the first property) is given by Eq.(\ref{eq:Bl}) since ${\partial E_i}/{\partial \cBB_{\vii}^{(1)}}=  \BBb_{i\vii}$.  
For the right half of the $\hat{B}_{\mu}$ matrix (the second property), ${\partial E_i}/{\partial \cBB_{\vii}^{(2)}} = \BBb_{i\vii} / \left(2 \sqrt{\ace_i^{(2)}} \right) $ by following the chain rule.
Extrapolation grades obtained with both linear, Eq.(\ref{eq:B_linear}), and non-linear, Eq.(\ref{eq:Bnl}), methods are compared in Sec.~\ref{sec:lnl} below.

\section{Comparison of uncertainty indication methods}
\label{sec:application}

In this section, we compare the performance of D-optimality and ensemble learning for uncertainty indication in both structure and composition space. We start with the structural extrapolation in elemental Cu, followed by composition extrapolation in binary Al-Ni. We use the term  extrapolation indicators to refer to both the extrapolation grade for D-optimality and the maximum deviation for ensemble learning.

\subsection{Structural extrapolation for copper}

The structural extrapolation for Cu was carried out for the Cu-II and Cu-III datasets introduced in Ref.~\cite{Bochkarev2022}. For the Cu-II set, we separate the analysis for the train and test subsets since they do not overlap.  The Cu-II dataset contains one thousand structures with energy up to 1\,eV above the fcc ground state. The Cu-III dataset is ten times larger and covers a much greater diversity of structures in a wider energy range up to the dissociation limit. Details of the Cu ACE potential can be found in Appendix \ref{app:ACEdetails}.

Figure~\ref{fig:Cu_mr_maxvol_gamma_hr_ensemble_dE_dF} shows errors in energy and forces versus the extrapolation indicators for all three datasets. In both methods, there exists a correlation between the indicators and the force and energy errors. For the Cu-II/train dataset, the extrapolation grade does not exceed one  by construction. For the Cu-II/test dataset, the majority of structures has comparably small energy and forces errors as well as the extrapolation indicators. As expected, significantly larger values of the extrapolation indicators and the corresponding errors are obtained for the more diverse Cu-III dataset. 

It is important to note that we do not observe structures with large errors for small extrapolation indicators. This means that there are no false-negative (FN) predictions and both indicators reliably predict structural extrapolation. Clearly, the extrapolation indicators cannot be used for estimation of absolute values of energies or forces, as very different structures may show similar energies or forces. 
Therefore, some structures with small errors display large extrapolation indicators. These cases correspond to false-positive (FP) predictions. This implies that the extrapolation indicators provide an upper bound for the expected errors in energy and forces.
\begin{figure*}
    \centering
    \subfloat[\centering D-optimality]{
        
     \includegraphics[width=0.82\columnwidth]{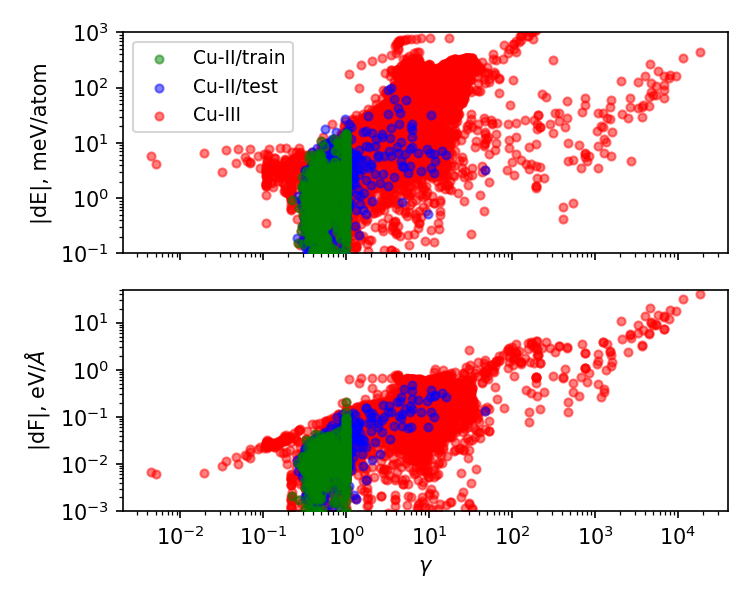}
        
    }
    \subfloat[\centering Ensemble learning]{
    
    \includegraphics[width=0.82\columnwidth]{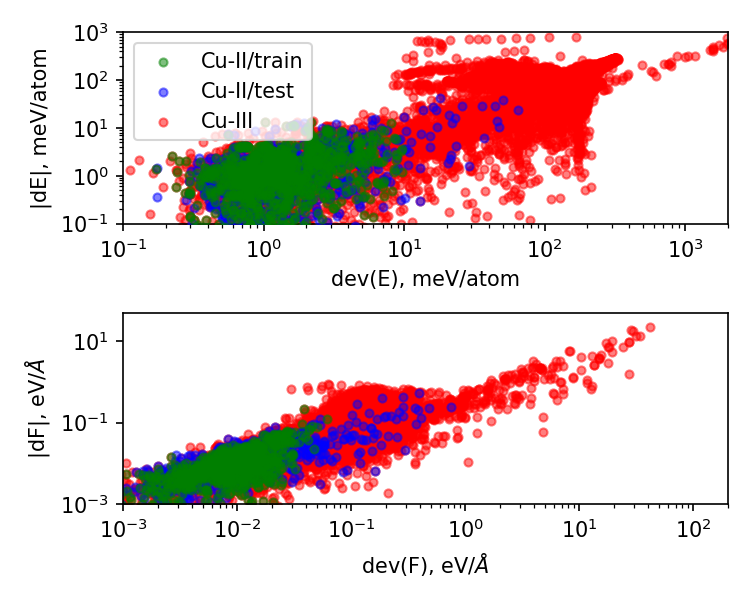}
    
    }
    \caption{Absolute energy and maximum force errors versus the extrapolation indicators for
    (a) the extrapolation grade in D-optimality and (b) the maximal deviation in ensemble learning for the Cu datasets.}
    \label{fig:Cu_mr_maxvol_gamma_hr_ensemble_dE_dF}
\end{figure*}

\begin{figure*}
    \centering
    \subfloat{
        \includegraphics[width=0.45\linewidth]{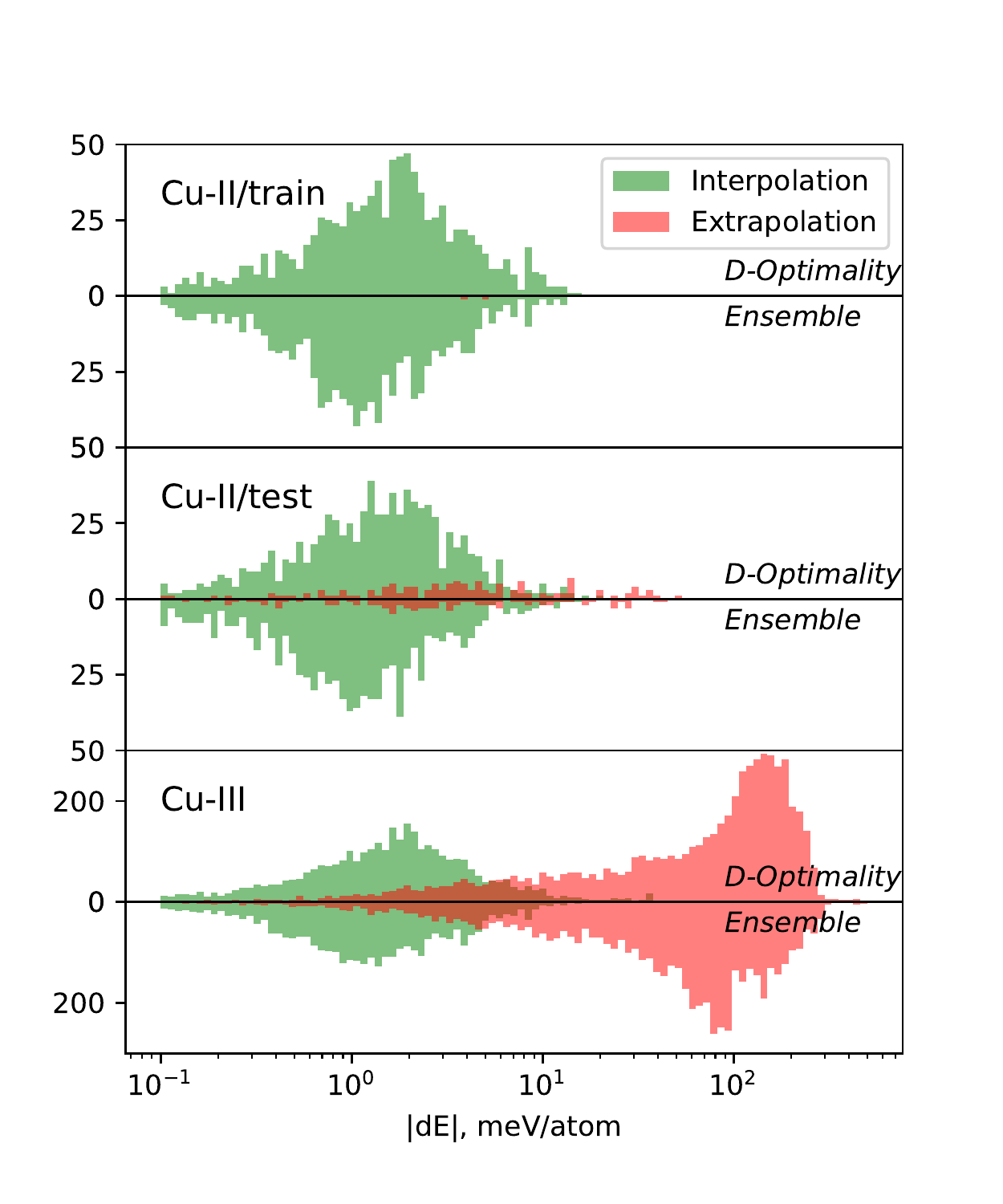}
    }
    \subfloat{
        \includegraphics[width=0.45\linewidth]{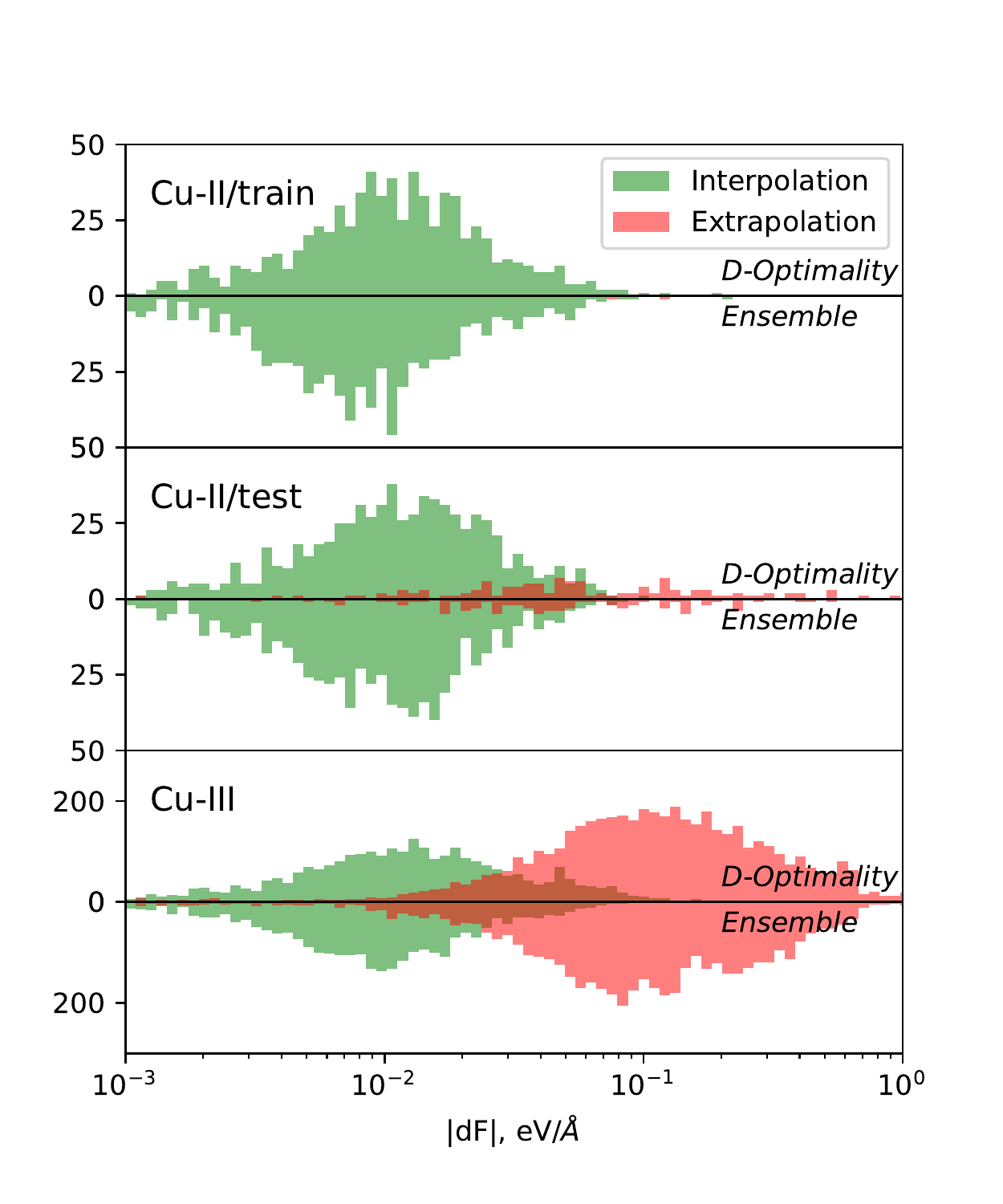}
    }
    \caption{Absolute energy error (left) and maximum absolute force error (right) counts for the  Cu datasets obtained using the D-optimality and ensemble learning criteria. See text for details.
    }
    \label{fig:Cu_mr_maxvol_mr_ensemble_dE_dist_dF_dist}
\end{figure*}

Figure~\ref{fig:Cu_mr_maxvol_mr_ensemble_dE_dist_dF_dist} shows the number of structures with a particular force or energy error according to the two criteria. For structures marked in green, the extrapolation indicators are below the threshold so that these structures are assumed to be in the interpolation regime. For structures marked in red, the indicators are above the threshold, suggesting extrapolation. The results from the D-optimality and ensemble learning analysis are consistent.

All structures in the Cu-II/train dataset are in the interpolating regime by construction. 
It is remarkable that for the Cu-II/test and Cu-III datasets, the interpolated structures show similar errors and error distributions as those for the Cu-II/train dataset. 
Structures with larger errors are all identified as extrapolated. The false positive (FP) prediction, where an extrapolation is indicated despite small errors, is seen for a small fraction of structures only. 
As mentioned above, this is due to the fact that very different structures can still have comparable energies.

In Appendix~\ref{app:CuExamples} we provide further illustrative examples where the extrapolation indicators are compared for a homogeneous deformation of bulk fcc crystal and a displacement of a single atom in bulk fcc crystal.

\subsection{Linear and non-linear extrapolation grades \label{sec:lnl}}

\begin{figure}
    \centering
    \includegraphics[width=1\columnwidth]{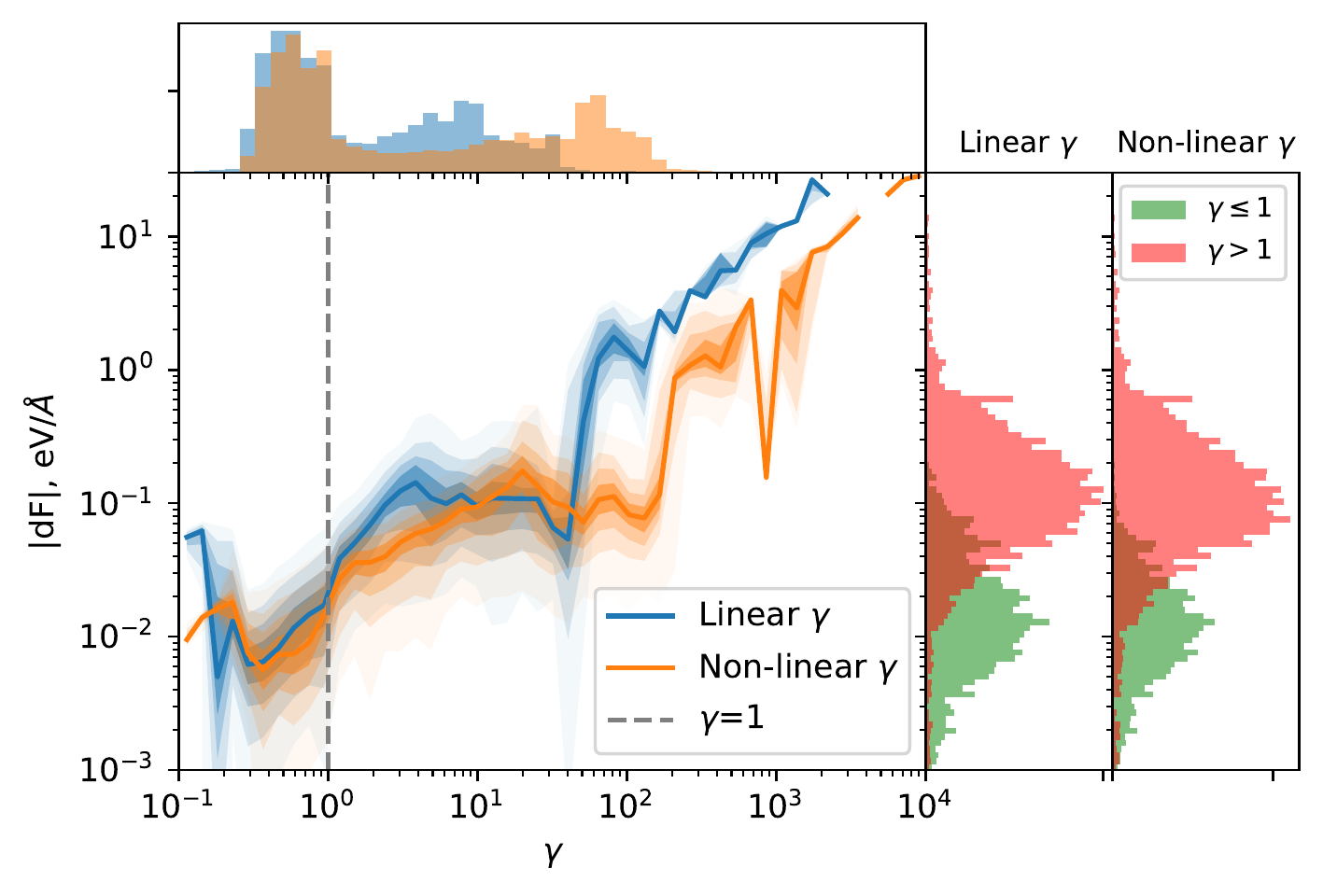}
    \caption{Analysis of force errors for the linear and non-linear extrapolation grades for the Cu-III dataset.
    }
    \label{fig:Cu3-full-linear-gamma-distr}
\end{figure}

The extrapolation grades obtained by the linear D-optimality criterion [cf. Eq.(\ref{eq:B_linear})] and its non-linear extension [cf. Eq.(\ref{eq:Bnl})] are compared in Fig.~\ref{fig:Cu3-full-linear-gamma-distr}. The comparison is done using an ACE parametrization that was trained on the Cu-II/train set but the test is carried out for the Cu-III set.

The solid lines show the median value of $|dF|$ as a function of $\gamma$ while the shaded regions of different transparency indicate the $\pm10\%$ -- $\pm40\%$ quantile ranges around the median. The upper histogram shows the distribution of $\gamma$ for both approaches. The two vertical histograms on the right panels display the force errors split into interpolation ($\gamma \leq 1$) and extrapolation ($\gamma > 1$).

\begin{figure}[h!]
    \centering
    \includegraphics[width=1.0\columnwidth]{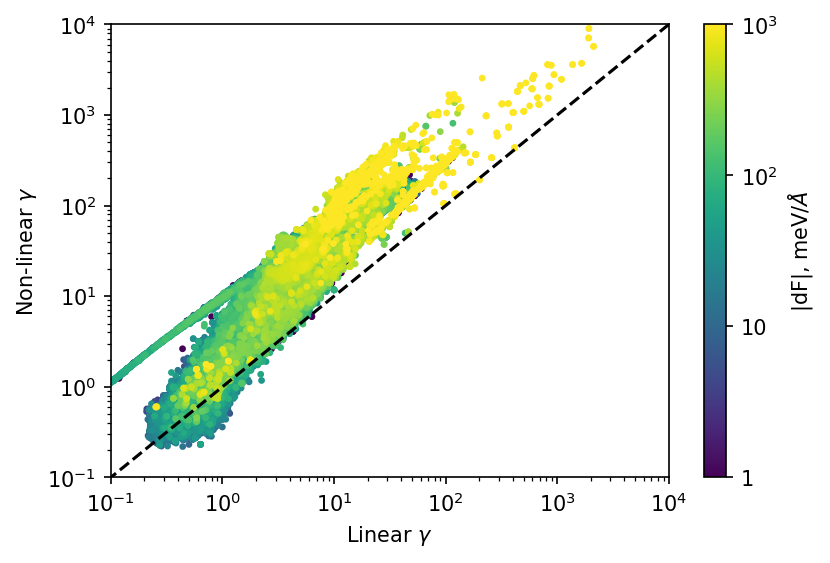}
    \caption{A correlation between the linear and non-linear extrapolation grades for the Cu-III dataset.
    }
    \label{fig:Cu3-full-linear-gamma}
\end{figure}

Overall, the linear and non-linear implementations show very similar behavior.
In particular, the division of the data into the interpolating and extrapolating regimes is nearly identical. 
Furthermore, errors remain reasonably small for extrapolating grades of the order of 10, which implies that the ACE model may have a large extrapolation radius and extrapolation grades larger than one can be tolerated in simulations. 
The correlation plot between the linear and non-linear extrapolation grades shown in Fig.~\ref{fig:Cu3-full-linear-gamma} reveals that the latter measure is more conservative, especially for values larger than ten.

This analysis confirms that both linear and non-linear extrapolation grades provide consistent predictions.  Due to its significantly lower computational costs and less conservative nature, we use the linear extrapolation grade in the following. Nevertheless, both approaches are included in the ACE implementation.

\subsection{Extrapolation in composition space for Al-Ni}

\begin{figure*}
    \centering
    \includegraphics[width=1\columnwidth]{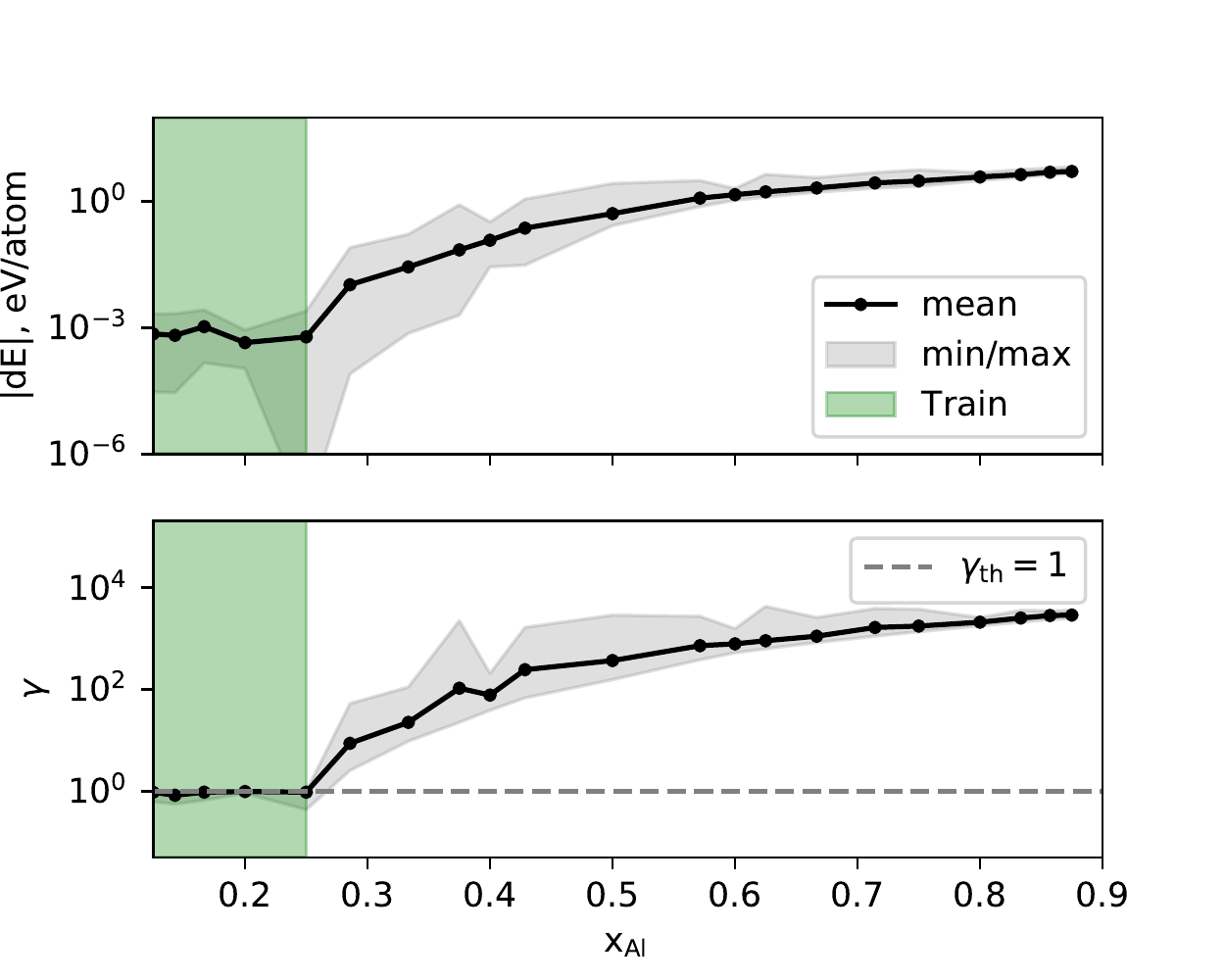}
    \includegraphics[width=1\columnwidth]{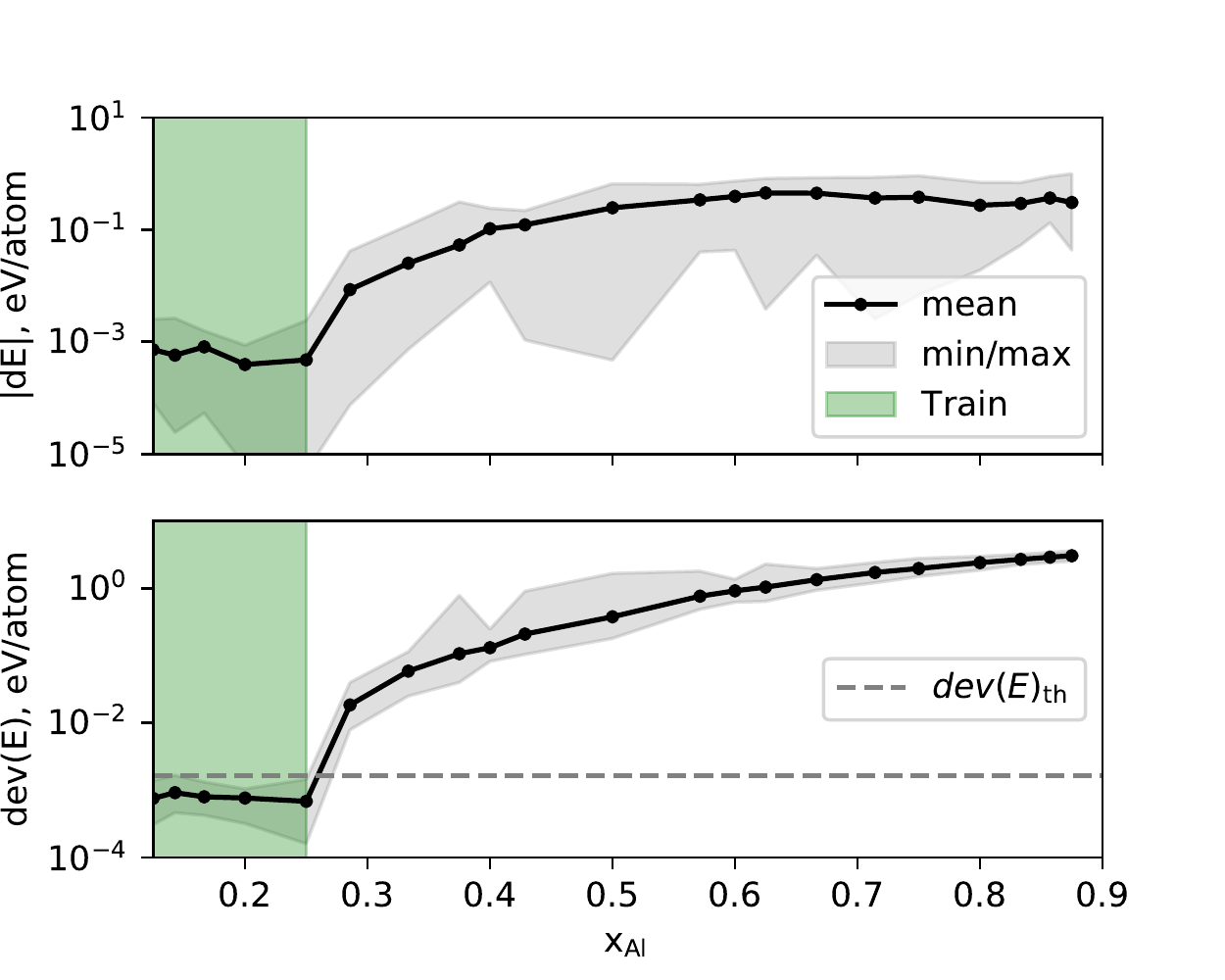}
    \caption{Compositional extrapolation for the Al-Ni binary dataset~\cite{nyshadham2019machine} using the D-optimality (left) and ensemble learning (right). All ACE models were fitted to structures with Al concentration $x_\textrm{Al} \le 0.25$. Dashed gray lines indicate corresponding extrapolation thresholds.
    }
    \label{fig:AlNi_Al25_dE_gamma}
\end{figure*}

The efficiency of extrapolation indicators in composition space was analysed for the binary Al-Ni system. We used a subset of the dataset 10B taken from Ref.~\onlinecite{nyshadham2019machine} that covers all possible decorations of binary fcc, bcc and hcp structures with up to eight atoms in the unit cell. Structures with Al concentration $x_\textrm{Al}\le 0.25$ were used for training and the remaining structures for testing. Figure~\ref{fig:AlNi_Al25_dE_gamma} shows the range of energy errors and the two extrapolation indicators, $\gamma$ and $\dev(E)$, as a function of Al content $x_\textrm{Al}$. As in the structure space, both uncertainty indication methods identify reliably the interpolation and extrapolation regimes also in the composition space.

\subsection{Summary of uncertainty indicators}

The examples presented above show that both uncertainty indicators perform equally well for structural as well as compositional extrapolation scenarios. However, from the practical point of view, the ensemble method has several drawbacks. Firstly, it requires training and evaluation of several models, which adds extra computational costs. Furthermore, the variation of the ensemble method may be limited by symmetry. For example, no force extrapolation can be established for high-symmetry crystal structures with vanishing forces.
In contrast, the calculation of the extrapolation grade $\gamma$ requires only vector-matrix multiplications that scale as $O(n_v^2)$, where $n_v$ is the number of basis functions per atomic species. 
A typical basis size in non-linear ACE models comprises of the order of $10^3$ functions per element, which makes the MaxVol algorithm computationally feasible and practical. 
The MaxVol algorithm also enables to systematically revise the active set of atomic configurations that improve the model transferability.
Therefore, we will focus in the following on the D-optimality criterion only and demonstrate its applicability in active learning scenarios.

\section{Active learning and sampling strategies}
\label{sec:AL}

In the context of data-driven development of interatomic potentials, active learning is a valuable autonomous procedure to improve model parameterizations in an iterative manner.
The active learning workflow consists of the following steps:
(1) generation of new atomic configurations, 
(2) sampling of configurations with high uncertainty using uncertainty indicators, and
(3) selection of the most representative configurations from the previous step.
These configurations are then reevaluated using the reference calculations and added to the training set. Finally, the model is retrained and applied until another uncertainty is detected. 

There exist various strategies for efficient generation of atomic configurations and identification of the unreliable cases.  
The most common way is to train ML potentials using preexisting collections of atomic configurations, molecular dynamics (MD) trajectories, or specifically generated configurations that are targeted at certain properties of interest (e.g., elastic, vibrational or defect properties). 
While sufficient for some applications, it is not the most efficient way and inherently restricts the portion of sampled configuration space.
A much more convenient and automatized approach is to explore the configuration space based on the uncertainty indicators.
Configurations with the extrapolation indicator above a certain threshold are selected for training and the accuracy and transferability of the model can be therefore continuously improved. 
In this section, we present several examples of different AL strategies for ACE.

\subsection{Comparison of selection strategies}

To examine the efficiency of different strategies for selection of structures from the pool of the extrapolative configurations, we retrained the ACE potential constructed on the small Cu-II/train dataset on additional structures selected in different ways from the large Cu-III/train dataset. 
The additional structures were selected based on (i) random sampling, (ii) the CUR decomposition~\cite{mahoney2009cur,imbalzano2018automatic}, and (iii) the D-optimality criterion using first the MaxVol algorithm followed by an addition of randomly selected structures.

\begin{figure}
    \centering
    \includegraphics[width=1.05\columnwidth]{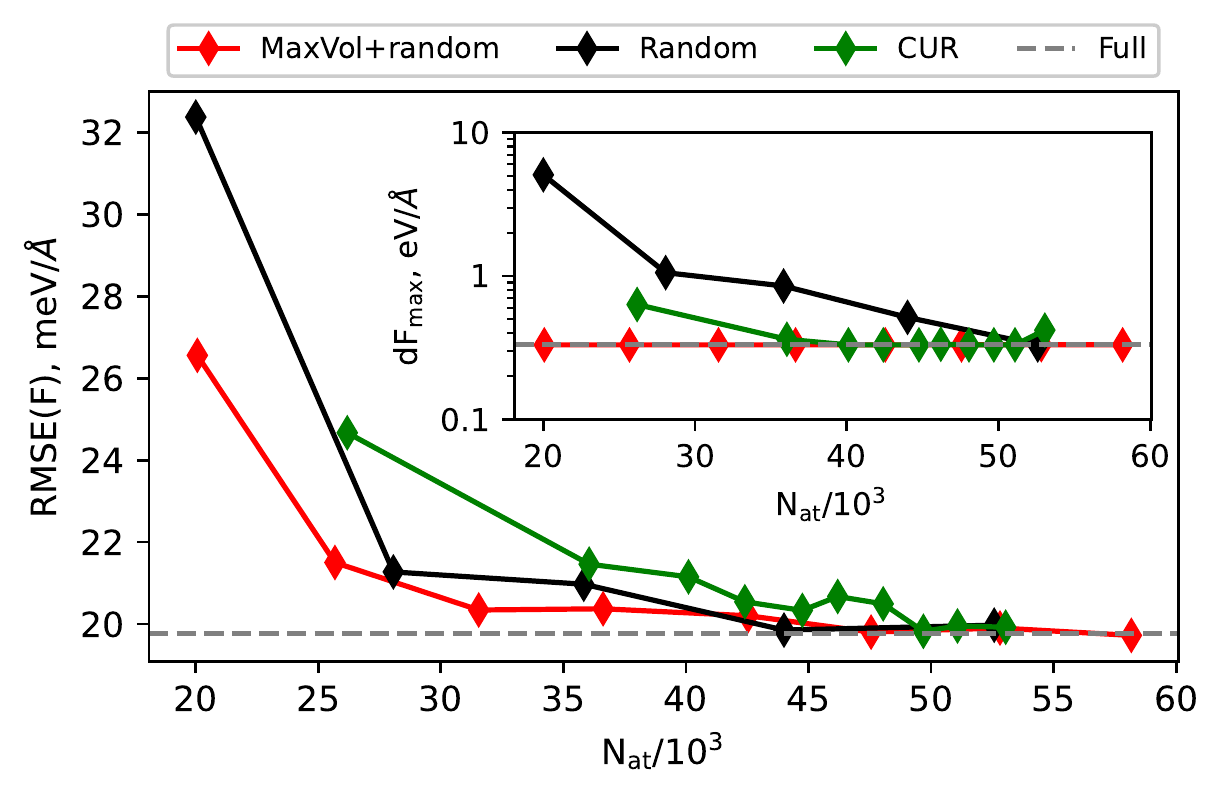}
    \caption{Force RMSE and maximum force error (inset) as a function of the training dataset size for three structure selection strategies; the dashed grey line mark results corresponding to the model trained on the full Cu-III/train dataset.
    }
    \label{fig:sample_selection}
\end{figure}

In Fig.~\ref{fig:sample_selection}, we compare the force RMSE as well as the maximum force error $dF_\mathrm{max}$ for the three selection strategies as a function of the total number of atoms $N_\mathrm{at}$ in the selected training set. The CUR decomposition shows the slowest convergence of force RMSE and a moderate maximum force error. The random selection leads to a slightly faster convergence of the force error but largest force error maxima, which implies large extrapolation and a significant lack of transferability. The MaxVol strategy results in both the most efficient learning and negligible maximum force errors during the whole retraining. In fact, the MaxVol retrained parametrization reaches errors comparable to those of the parametrization trained on the complete Cu-III/train dataset.
The comparably poor performance of CUR may be attributed to the very large variation of the values of the basis functions $\BBb_{i \vii}$, that span eight orders of magnitude in the training set.

\subsection{Molecular dynamics of water}

MD simulations can probe large portions of the configuration space and therefore can be conveniently used for structure generation and exploration. We investigated the training of ACE for water using  the whole AL loop based on MD trajectories. The reference data were generated using DFT, as implemented in VASP\,5.4~\cite{kresse1993ab,kresse1996efficiency,kresse1996efficient}. We employed the projector augmented wave (PAW) method~\cite{kresse1999ultrasoft} and the revPBE exchange correlation functional~\cite{zhang1998comment} with the DFT-D3 correction~\cite{grimme2010consistent}. A plane wave cutoff energy of $E_\mathrm{cut}=400$\,eV and a single $\Gamma$-point in the Brillouin zone with Gaussian smearing of width $\sigma=0.01$\,eV were found sufficient to obtain well converged values of energies and forces.

\begin{figure}
    \centering
    \includegraphics[width=1.0\columnwidth]{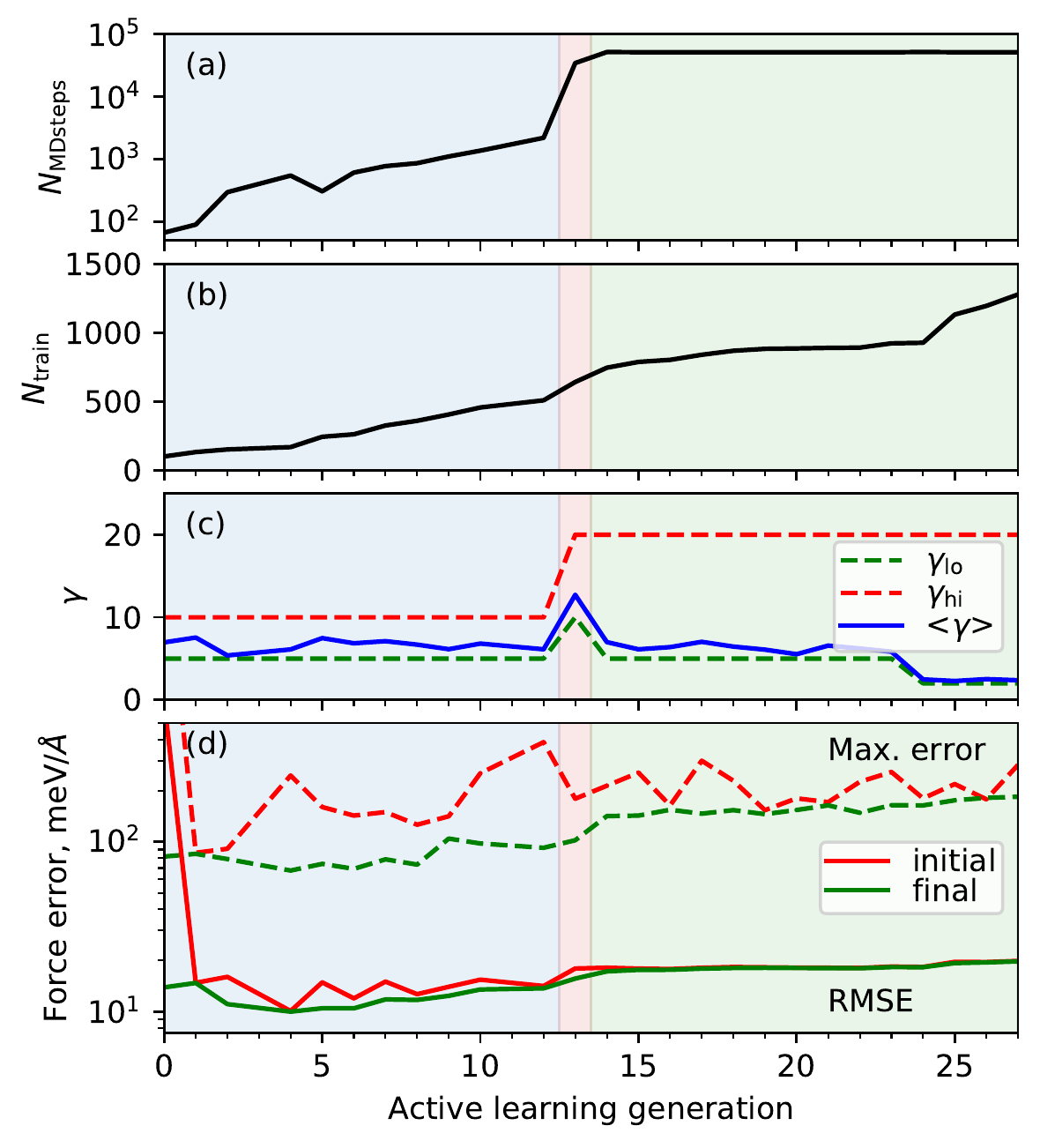}
    \caption{Analysis of AL procedure for water: 
    (a) number of successful MD steps before extrapolation grade exceeds $\gamma_\mathrm{hi}$,
    (b) training set size, 
    (c) extrapolation grade range [$\gamma_\mathrm{lo}$,
    $\gamma_\mathrm{hi}$] and average grade $\langle \gamma \rangle$, 
    (d) initial and final (before and after ACE model retraining) force error metrics (RMSE and max. error).
    Three highlighted regions on each plot mark MD-exploration regimes, see text for more details.
    }
    \label{fig:water_al_hist}
\end{figure}

The initial ACE parametrization was trained on a dataset consisting of a short DFT MD trajectory with 100 steps at 300 K using a supercell with 64 water molecules and density $\rho=0.9995$\,g/cm$^3$. The subsequent training was done using AL, based on a series of longer NVT MD simulations at 300\,K for up to 50,000 steps (covering 25~ps with a timestep 0.5~fs) using \LMPS~\cite{LAMMPS}. Each of these simulations always started from the  same initial configuration.
The extrapolation grade $\gamma$ was computed every $n_\gamma$ steps and the atomic configuration stored if  $\gamma_\mathrm{lo} \le \gamma \le \gamma_\mathrm{hi}$ (see below). If the extrapolation grade during the MD run exceeded $\gamma_\mathrm{hi}$ the simulation was stopped. 
After each simulation, the MaxVol algorithm was used to select from the stored configurations those that provide the widest extension of the current active set.
These structures were computed using VASP, added to the training set, and the next ACE generation was retrained. 

The analysis of crucial parameters throughout the whole AL series, consisting of 27 generations of ACE models, is shown in Fig.~\ref{fig:water_al_hist}.
The blue, red and green shadings in all panels mark three regimes we examined.
In the first 12 MD runs (blue shading), the extrapolation grade was evaluated at every step ($n_\gamma=1$), and the configurations were selected for a narrow range between $\gamma_\mathrm{lo}=5$ and $\gamma_\mathrm{hi}=10$. 
As can be seen on the top panel in Fig.~\ref{fig:water_al_hist}(a), none of the MD runs reached the 50,000 steps and all were interrupted much sooner due to the extrapolation grade exceeding $\gamma_\mathrm{hi}$. 
During the 13$^{th}$ MD run (red shading), we boosted the sampling of more uncertain configurations by increasing $n_\gamma=10$, and the range of $\gamma_\mathrm{lo}=10$ and $\gamma_\mathrm{hi}=20$. 
Finally, during the last 14 MD runs (green shading), the range of $\gamma$ was further increase by decreasing the lower threshold to $\gamma_\mathrm{lo}$ to 5 and eventually 2. 
During the final regime, all MD simulations always reached the targeted 50,000 steps, see Fig.~\ref{fig:water_al_hist}(a). 

Figure~\ref{fig:water_al_hist}(b) shows that the increase of the number of structures in the training dataset during the whole AL series follows approximately a linear trend.  
The final size of the training dataset was 1277. 
Figure~\ref{fig:water_al_hist}(c) depicts both the minimum $\gamma_\mathrm{lo}$ and maximum $\gamma_\mathrm{hi}$ ranges as well as the average extrapolation grade $\langle \gamma \rangle$.

Figure~\ref{fig:water_al_hist}(d) shows the dependence of the initial and final force errors during the AL series. Both the final force RMSE and the maximum error somewhat increase due to the increase of the training set size.  It is interesting to note that the initial maximum force error, which is computed after adding newly selected structures from the previous generation but before the new potential is trained, is significantly larger than the final one. This confirms that the newly included structures are described very poorly by the previous generation. In contrast, the initial and final force RMSE values change very little. Thus, the main effect of the AL procedure is to discover outliers and reduce their errors.

\begin{figure}
    \centering
    \includegraphics[width=1.0\columnwidth]{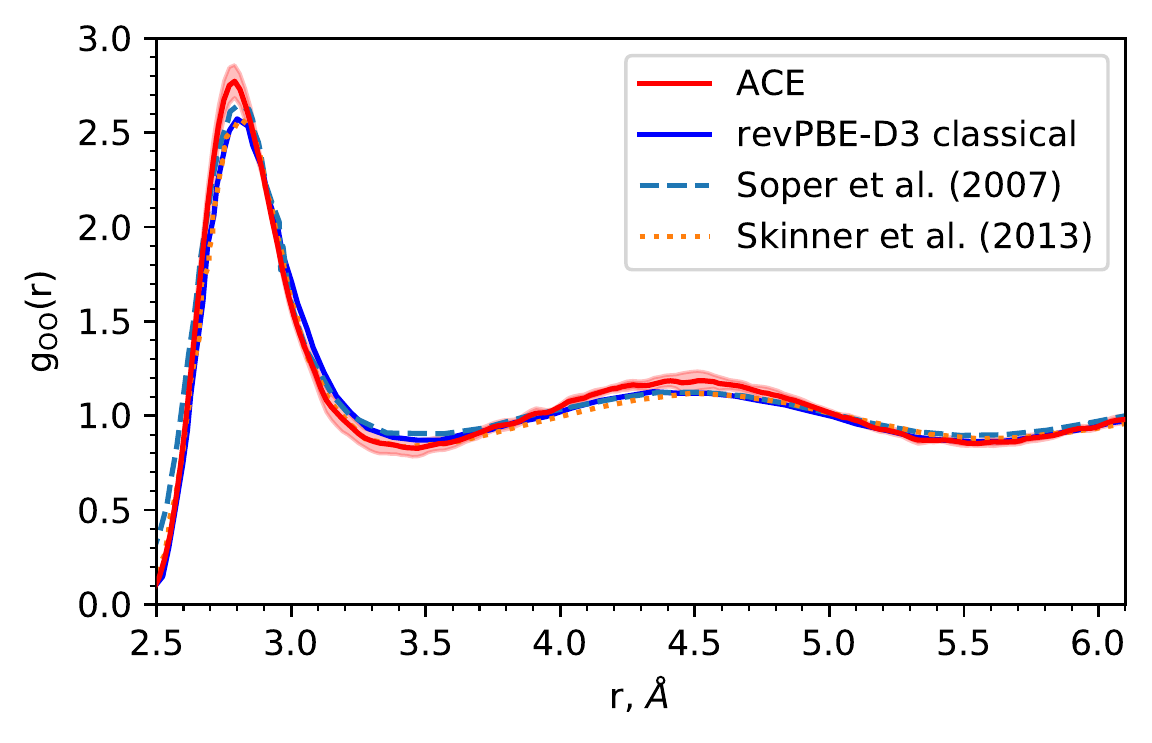}
    \caption{Radial distribution function for O-O in water. Comparison of ACE, DFT~\cite{marsalek2017quantum} and two experiments~\cite{soper2007joint,skinner2013benchmark}. The shaded region marks standard deviation from five independent MD runs.}
    \label{fig:H2O_OO_rdf}
\end{figure}

Using the final ACE parametrization, we computed the oxygen radial distribution function (RDF) $g_\mathrm{OO}(r)$ in five independent MD runs. The results are shown in Fig.~\ref{fig:H2O_OO_rdf} together with results from ab initio MD simulations~\cite{marsalek2017quantum} and two experiments~\cite{soper2007joint, skinner2013benchmark}.

\subsection{Active exploration of Li$_4$ clusters}
\label{sec:gamma-li4-clusters}

Even though MD simulations are straightforward for sampling of atomic configurations, they may be computationally too demanding or impractical for certain configurations (e.g. extended defects, rare events sampling). Another possibility, often more convenient and effective, is to generate the structures using the uncertainty indicators. In the context of ACE, we term this procedure as  active exploration (AE). Similar strategies based on gradients of uncertainty indicators have been proposed recently~\cite{van2022hyperactive,kulichenko2022uncertainty}. 
Different from these approaches, AE in ACE does not require  retraining of ACE parametrization during the buildup of the training dataset, simply because the active set can be updated without changing the potential. 

We demonstrate the functionality of AE for 4-atom Li clusters, which are described by only six coordinates.
This enables to perform an unbiased assessment of the AE performance by comparing it to brute-force sampling. The ACE AE implementation proceeds as follows:
\begin{enumerate}
    \item Start from a small reference dataset of structures which may be created from a short MD trajectory or an existing database.
    \item Train an ACE parametrization and determine the active set.
    \item Randomly select a structure from the reference dataset and modify its atomic positions to maximize the uncertainty, which is measured as the maximum extrapolation grade of any atom in the structure.
    Once the uncertainty  exceeds a chosen maximum threshold value $\gamma_\mathrm{hi}$, maximization process terminates.
    Certain geometrical constraints can be applied at this stage.
    \item Add the new structure to the active set. 
    \item Repeat the steps 3 and 4 until the number of new structures does not exceed $\Delta N_\mathrm{struct}$  or the maximum number of optimization attempts is reached.
    \item Compute energies and forces of the collected structures with DFT and update the training dataset.
    \item Repeat from the step 2. until no structures with large extrapolation grade can be found.
\end{enumerate}

For the Li clusters, we used $\gamma_\mathrm{hi} = 100$ and up to $\Delta N_\mathrm{struct}=20$ structures to be collected in one iteration before the reference dataset was extended. The initial training set contained 1351 Li$_2$ and Li$_3$ clusters without any Li$_4$ clusters. The modification of atomic positions in the distorted structure (point 3) was carried out using the Nelder-Mead method, which does not require the knowledge of gradients of atomic positions with respect to the extrapolation grade. The validation was done for a test set containing 10816 Li$_4$ clusters with almost uniform sampling of the 6-dimensional configuration space.

\begin{figure}
    \centering
    \includegraphics[width=1.0\columnwidth]{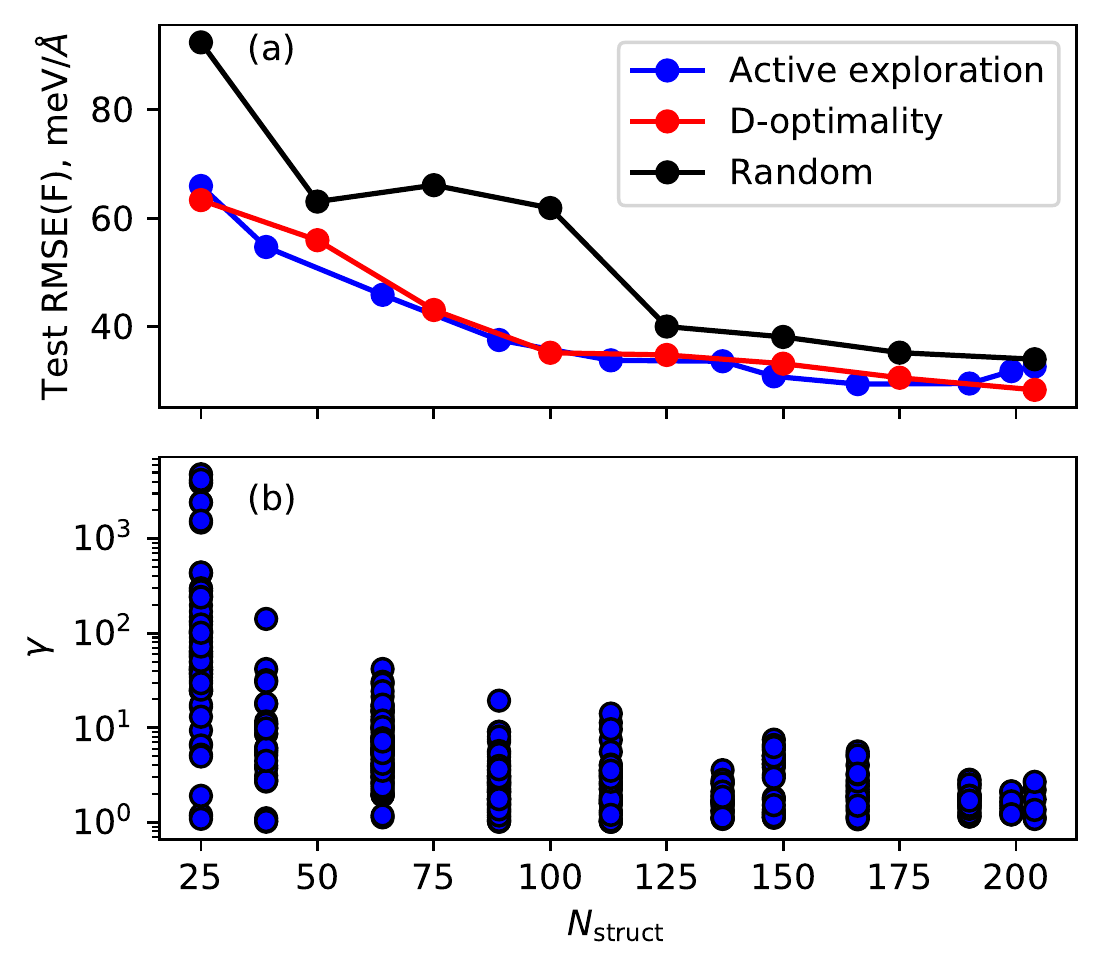}
    \caption{Comparison of active exploration, D-optimality and random selection strategies for training of Li$_4$ clusters (a), and variation of the extrapolation grade during active exploration (b), see text for more details.}
    \label{fig:Li4_exploration}
\end{figure}

We compared AE to a random selection from the complete test set and to subsets of increasing size selected using the MaxVol D-optimality criterion. Learning curves from all three approaches are plotted in Fig.~\ref{fig:Li4_exploration}(a). The AE and D-optimality selection perform equally well while the random selection shows the worst performance and the slowest convergence among the three methods. 

Figure~\ref{fig:Li4_exploration}(b) shows the range of $\gamma$ values for the new structures collected in each AE iteration.
The maximum extrapolation grade in each iteration effectively characterizes a relative change of the hypervolume of the active set after a new atomic environment is added.
Initially, $\gamma$ reaches very large values ($> 10^3$) since the initial training set encompasses only a small part of the configuration space.
In the later stages, the maximum $\gamma$ decreases exponentially because it is more and more difficult to find new atomic environments with large extrapolation grades. 
Nevertheless, the configuration space remains to be actively explored and the hypervolume of the corresponding active set keeps expanding.
It should be also noted that each newly added configuration helps to steer the search of next configurations.

The AE approach is convenient for exploration of local atomic distortions involving few atoms (e.g., migration pathways in a larger supercell, structures of defects, adsorption of molecules at surfaces, dissociation of molecules, and other rare events) since its efficiency decreases with increasing number of atoms due to the quasi-stochastic optimization.
However, it is possible to incorporate several locally explored atomic distortions into one supercell to increase the computational efficiency.

\subsection{Active learning for local atomic environments}

One of common problems encountered in AL schemes is how to efficiently perform the procedure when the region of large extrapolation encompasses just a few atoms in otherwise large simulation block. Large extrapolation grades often occur in the cores of extended defects (dislocations, grain boundaries, crack tips, etc.) embedded in large bulk environment containing many hundreds or thousands of atoms. Such structures are impossible to compute with DFT methods and add them to the training dataset. 

The simplest possibility is to just cut out a small, free-standing cluster about the atom with large $\gamma$. However, apart from dealing with proper surface terminations and relaxations, this usually leads to marked changes of the electronic structure so that the original bonding environment is not preserved. A viable option in some cases (usually in metals) is to cut out instead a suitable rectangular region and impose periodic boundary conditions. Atoms in the outer shell (typically outside the cutoff radius of the potential) are then allowed to relax to minimize their $\gamma$, while atoms within the cutoff radius of the atom(s) of interest with large $\gamma$ are kept fixed. 

\begin{figure}
    \centering
    \includegraphics[width=1.0\columnwidth]{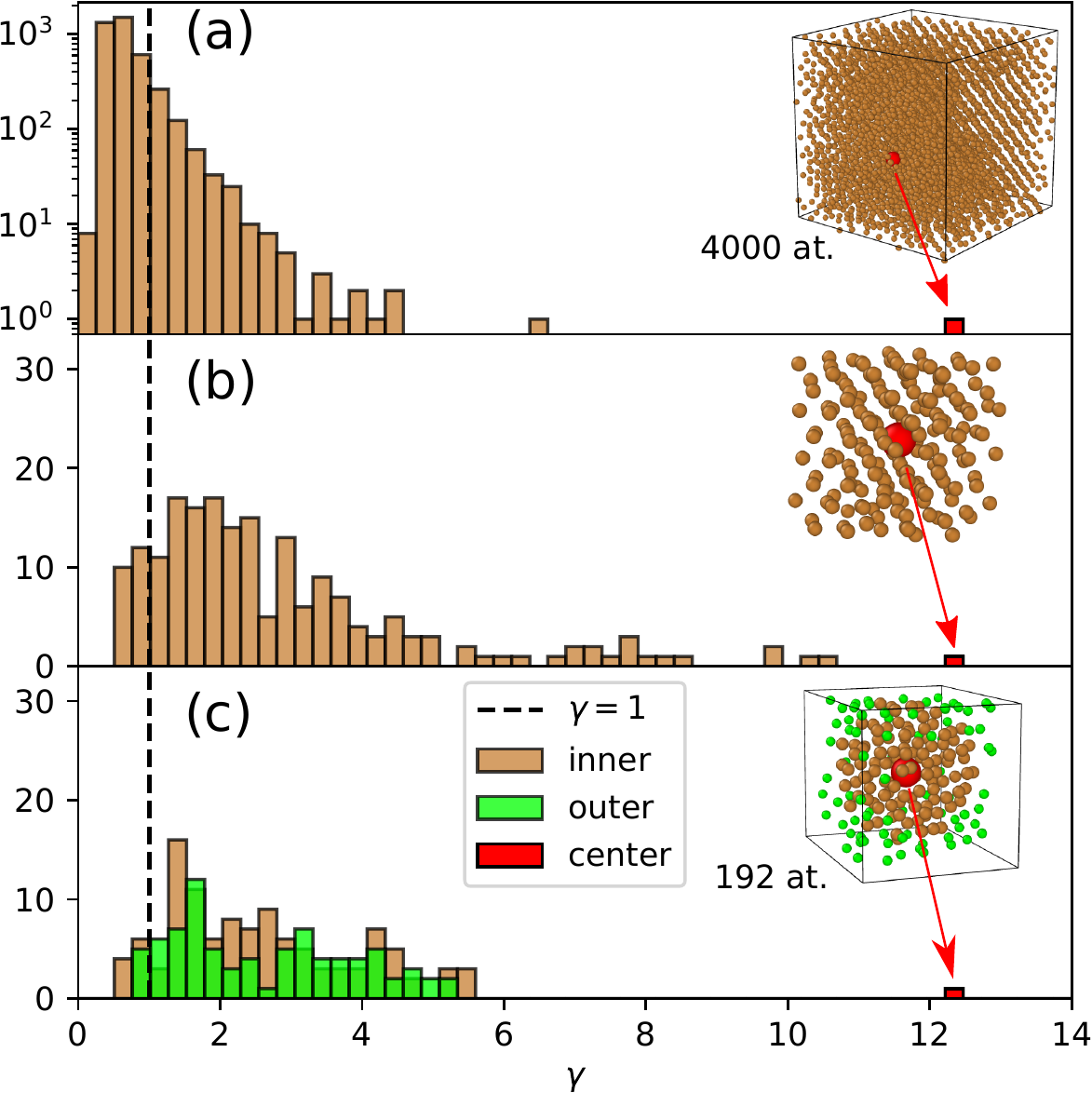}
    \caption{ Distribution of extrapolation grade in Cu-fcc system: (a) original large cell containing 4000 atoms, (b) an isolated cut-out cluster around the red atom with large $\gamma$; (c) the same cell as in (b) but with periodic conditions and optimized atomic positions of outer shell atoms (green atoms).}
    \label{fig:Cu_clust_pbc}
\end{figure}

We investigated this procedure for a large supercell in Cu containing 4000 atoms. Figure~\ref{fig:Cu_clust_pbc}(a) displays the distribution of atomic extrapolation grades for an MD snapshot at T=1400\,K. 
The red atom has the largest extrapolation grade greater than 12. Figure~\ref{fig:Cu_clust_pbc}(b) shows the results for a small periodic cell with free surfaces that was cut out about this atom. As expected, the presence of free surfaces leads to a significant increase of the extrapolation grades on some atoms. When we apply periodic boundary conditions and relax the outer atoms, all atoms show moderate extrapolation grades except for the central atom of interest, as shown in Fig.~\ref{fig:Cu_clust_pbc}(c).

Even though not universally applicable, this procedure allows to extract a specific atomic environment from large scale MD-simulations that is suitable for DFT calculation and thus opens the road towards the learning on-the-fly in very large systems.

\section{Conclusion}
\label{sec:conclusion}
In this work, we compared the performance of two approaches for uncertainty indication of ACE models based on the D-optimality criterion and ensemble learning. 
While both approaches show comparable predictions, the extrapolation grade based on D-optimality and the MaxVol algorithm is more computationally efficient since it does not require  training and inference of multiple models.

Both linear and non-linear extrapolation grades serve as reliable uncertainty indicators in various practical cases, including structural and compositional extrapolation as well as active learning.
These uncertainty indicators enable active exploration of new structures, which are unlikely to be encountered in molecular dynamics. 
This can be done without retraining of ACE models, opening the way to the automated discovery of rare-event configurations.
Finally, we demonstrated that the active learning is also applicable to explore local atomic environments from large-scale MD simulations.
In conclusion, some form of the uncertainty indication should be part of every interatomic potential to ascertain its reliability and transferabilty.

\section{Code availability}
The \PM code and the \texttt{\detokenize{tensorpotential}} fitting backend are available at \emph{https://github.com/ICAMS/python-ace} and \emph{https://github.com/ICAMS/tensorpotential}, respectively. Extrapolation grade calculation in \LMPS is described at \emph{https://docs.lammps.org/pair\_pace.html}.

\begin{acknowledgments}
RD acknowledges funding through the German Science Foundation (DFG), projects number 405621217 and 403582885. YL acknowledges funding through the German Science Foundation (DFG), project number 405602047. MM acknowledges funding through the German Science Foundation (DFG), project 405621081. RD acknowledges computational resources of the research center ZGH.
\end{acknowledgments}


\appendix

\section{Deviation of site energies\label{app:atomicdev}}

\begin{figure}[h!]
    \centering
    \includegraphics[width=1.0\columnwidth]{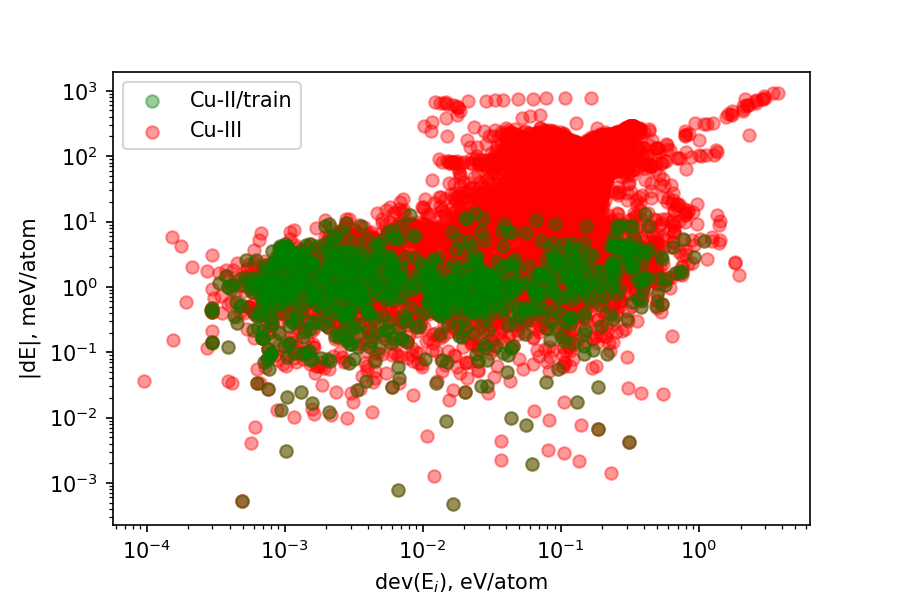}
    \caption{A scatter plot showing the maximum deviation of atomic energies $E_i$ and the absolute energy error for the corresponding structure $|dE|$.}
    \label{app:fig:Cu_hr_ensemble_dev_dE_i}
\end{figure}

As shown in Fig.~\ref{app:fig:Cu_hr_ensemble_dev_dE_i}, the maximum deviation of atomic energies $\dev(E_i)$ cannot be used for uncertainty indication because there exists no correlation between this quantity and the absolute energy error $|dE|$ for the complete structure.

\section{Generation of ensembles of ACE models
\label{app:ens_randomization}}

For ensemble generation, we employed two levels of randomization:
\begin{itemize}
\item[(i)] Weak randomization: The ACE radial functions (see Appendix C in Ref.~\cite{Bochkarev2022}) are optimized only once and kept fixed afterwards.
The basis function coefficients are initialized from a normal distribution $\cBB_{\mathbf{\mu n l L}}^{(p)}  \sim \mathcal{N}(0,\sigma^2) $. 
This approach speeds up the ensemble evaluation since it does not require recalculation of ACE basis functions for each member of the ensemble.
\item[(ii)] Strong randomization: The radial expansion coefficients are initialized as $c_{nlk}^{\mu_i \mu_j}=\delta_{nk} +\epsilon $, where $\epsilon \sim \mathcal{N}(0,\sigma^2)$. The basis function coefficients are initialized as in (i).
\end{itemize}
We set the standard deviation for all normal distributions to $\sigma = 10^{-4}$ and the number of potentials in each ensemble $N_\textrm{ens} = 5$. 

To determine the thresholds of total energy deviation  {$\mathrm{dev}(E)$ [Eq.(\ref{eq:devEF})]}, that separate interpolation and extrapolation, we use the following approach: 
we find the outlier energy error threshold $dE_\textrm{out}=Q_3 + 1.5\,\mathrm{IQR}$, where $Q_3$ is the third quartile of energy train error $dE$ distribution and $\mathrm{IQR} = (Q_3-Q_1)$ is the interquartile range. 
Then we select the maximum deviation threshold $\mathrm{dev}(E)_\textrm{th}$ for structures from the training set whose errors are below an outlier bound $dE_\textrm{out}$.
The deviation threshold for forces {$\mathrm{dev}(F_i)$} is computed in a similar manner.

\begin{figure*}
    \centering
    \subfloat[\centering]{
        \includegraphics[width=0.5\linewidth]{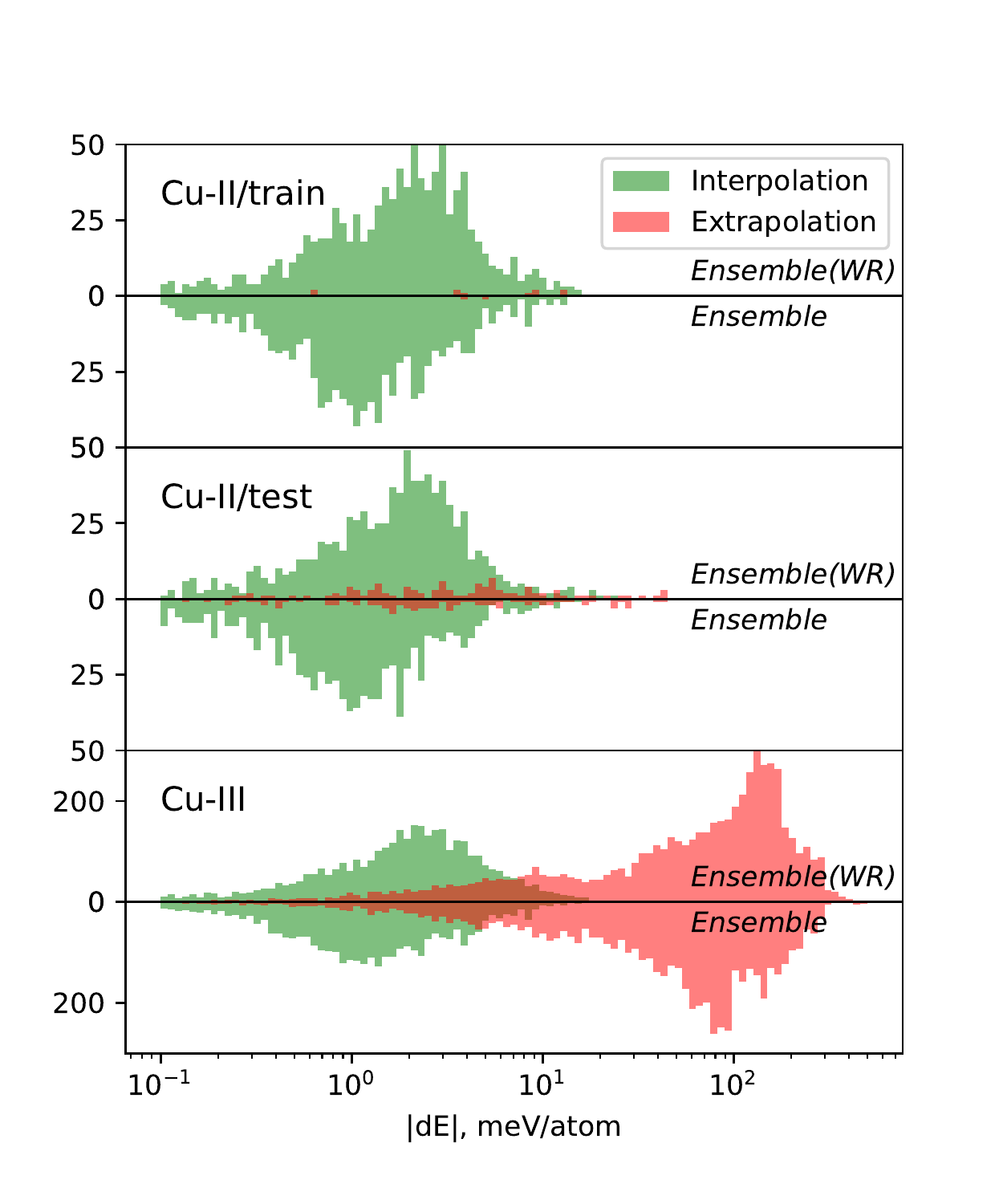}
    }
    \subfloat[\centering]{
       \includegraphics[width=0.5\linewidth]{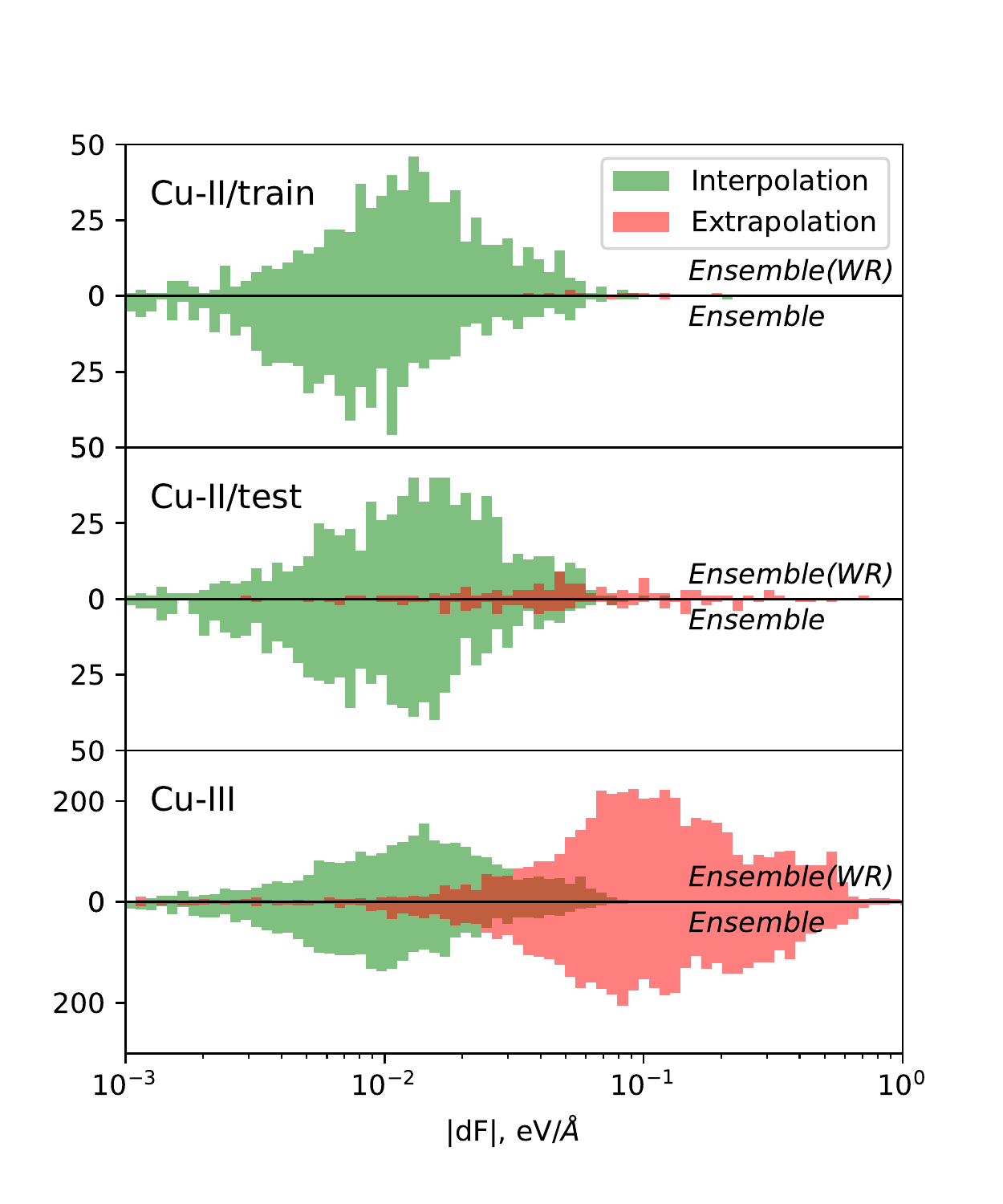}
    }
    \caption{Absolute energy error (a) and maximum absolute force error (b) counts for the  Cu datasets obtained using the  energy and forces deviation criteria for strongly and weakly randomized (WR) ensembles correspondingly. See text for details.}
    \label{app:fig:Cu_wr_ensemble_hr_ensemble_dE_dF_dist}
\end{figure*}

To achieve a better decorrelation, we use a strongly randomized ensemble. 
However, as demonstrated in Fig.~\ref{app:fig:Cu_wr_ensemble_hr_ensemble_dE_dF_dist}, a weakly randomized ensemble shows a very similar error distribution. 
The weak randomization is computationally more efficient since the B-basis functions are computed only once for the same radial functions. 
A similar approach for uncertainty indication was recently demonstrated for linear ACE models~\cite{van2022hyperactive}.

\section{Details of ACE potentials \label{app:ACEdetails}}

The configurations of ACE potentials used in this work are summarized in Table~\ref{tab:aceconfig}.

\begin{table*}[]
    \centering
    \caption{Summary of ACE configurations: cutoff radius ($r_c$), type of radial basis functions, $\nu$-order, n$_\mathrm{max}$, l$_\mathrm{max}$, and the maximum number of functions per element (\# func/elem) for each order $\nu$ for this configuration.}
    \begin{tabular}{ccccccc}
    \hline
    System & r$_c$ (\AA)& Radial basis function & $\nu$-order & n$_\mathrm{max}$ & l$_\mathrm{max}$ & \# func/elem \\ 
    \hline
    Cu  & 7.0 & SBessel & 7 & 15/5/4/3/3/2/1 & 0/3/2/2/1/1/1 & 15/60/160/300/144/50/3  \\
    Al-Ni  & 7.0 & SBessel & 5 & 13/3/2/2/1 & 0/2/2/1/1 & 21/51/55/21/2  \\
    Water & 7.0 & SBessel & 6 & 15/4/3/2/2/1 & 0/2/2/2/1/1 & 30/108/490/245/476/50  \\
    Li clusters & 7.0 & SBessel & 6 & 15/4/3/2/2/1 & 0/4/3/2/1/0 & 15/50/137/75/30/1  \\
    \hline
    \end{tabular}
    \label{tab:aceconfig}
\end{table*}

\section{Extrapolation indicators for simple geometries\label{app:CuExamples}}

\begin{figure*}[ht]
    \centering

    \includegraphics[width=0.45\linewidth]{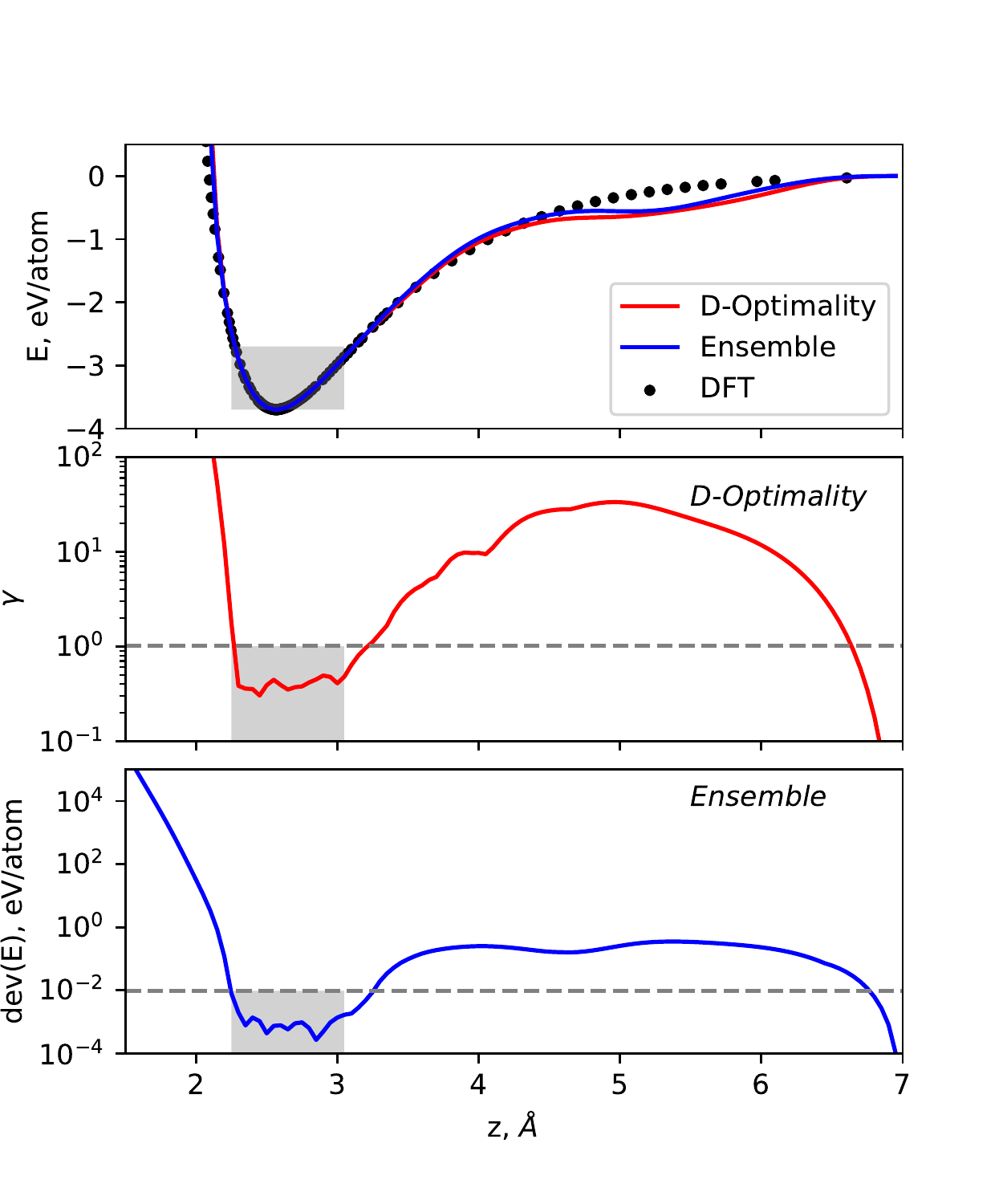}
    \includegraphics[width=0.45\linewidth]{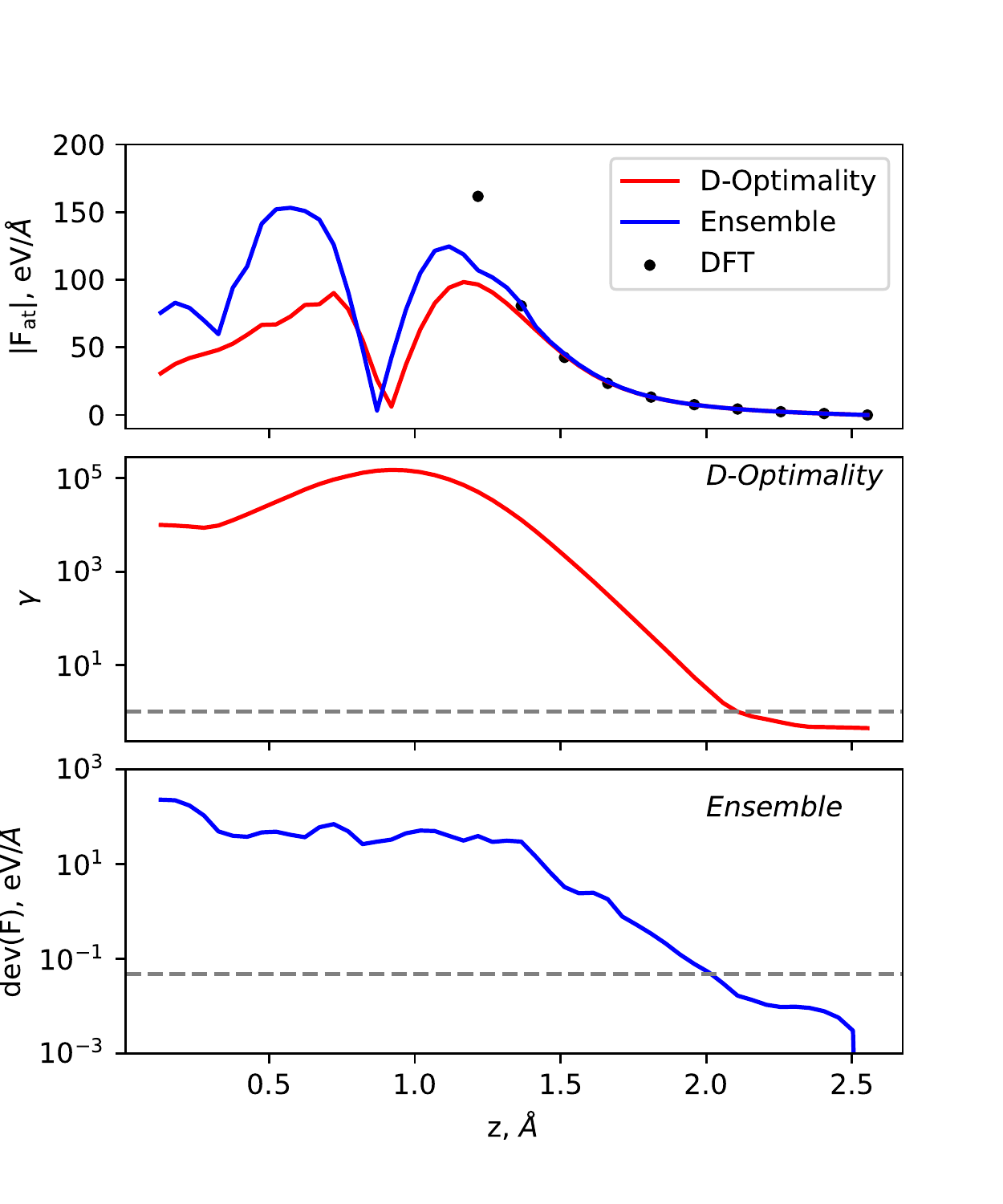}
     
    \caption{Extrapolation indicators for homogeneous deformation of fcc Cu (left) and displacement of a single atom in a fcc crystal (right). See text for details.}
    \label{fig:Cu_fcc_Ez_mr_maxvol_mr_ensemble}
\end{figure*}

We illustrate the extrapolation indicators (extrapolation grade and deviation of energies and forces) for the homogeneous deformation of a Cu-fcc crystal and the displacement of a single atom in a fcc crystal, displayed in the left and right column of Fig.~\ref{fig:Cu_fcc_Ez_mr_maxvol_mr_ensemble}, respectively.
The energy and volume ranges for training dataset are shown as gray shaded area. 
The predicted energies for large deviations are compared to reference data for a single model (D-optimality) and an ensemble. Below the extrapolation grade and the maximum deviation in energy is shown. As before, both extrapolation indicators give an accurate and similar demarcation of the interpolating and extrapolating regions.

In the right column, a Cu atom in fcc-supercell is approaching its neighbour starting from its equilibrium position, while all other atoms remain fixed. The interpolation indicators predict extrapolation for distances smaller than about 2.1\,\AA, while in fact accurate forces are predicted down to a distance of approx.~1.5\,\AA.
These findings again demonstrate that ACE models have a large extrapolation radius.

\section{Usage}
\label{app:implementation}

The extrapolation grade $\gamma$ is obtained from the active set matrix.  
The active set can be computed using a given ACE potential and its corresponding training set.
If the training set is too large to fit into the memory, an approximate active set can be computed iteratively by batches.

The inverted active set (ASI file) together with the corresponding ACE potential (YAML file) can be used in \LMPS as \emph{pace/extrapolation} pair style or in Python as ASE calculator~\cite{ase-paper}.
In \LMPS, the extrapolation grade $\gamma$ is computed every $N$ steps. 
The recursive evaluator \cite{lysogorskiy2021} is used for the steps in between to avoid more expensive matrix-vector multiplications at every step. 
In this way one preserves the control of the extrapolation grade, which is especially useful in large production simulations.
Simulations {can be} stopped when the extrapolation grade on any atom exceeds the threshold $\gamma_\mathrm{high}$. 
Optionally, structures for which at least one atom has an extrapolation grade above $\gamma_\mathrm{low}$ are written into a file. 
Typical values are $\gamma_\mathrm{low} \approx 2-10$  and $\gamma_\mathrm{high} \approx 10-25$ for linear extrapolation grade and can be larger for non-linear one.

\newpage
\bibliography{ace}

\end{document}